\documentclass[12pt]{iopart}

\usepackage{graphicx}
\usepackage{algorithm}
\usepackage{algpseudocode}
\begin{document}

\title[Utilizing rate-independent hysteresis for analog computing]{Utilizing rate-independent hysteresis for analog computing}
\author{Lina Jaurigue and Kathy Lüdge}

\address{Technische Universität Ilmenau}
\ead{lina.jaurigue@tu-ilmenau.de}
\vspace{10pt}
\begin{indented}
\item[]Jan 2025
\end{indented}

\begin{abstract}

Physical systems exhibiting hysteresis are increasingly being used in neuromorphic and in-memory computing research. Generally, the resistance switching of devices with rate-independent hysteresis are being investigated for their use as trainable weights in neural networks, whereas the dynamics of devices showing rate-dependent hysteresis are being investigate for their potential as nodes in, for example in reservoir computing systems. In our work we instead show the computing potential of a simple rate-independent hysteresis system. We show that by driving a system of only two linear branches with time-multiplexed inputs it is possible to generate nonlinear transforms and perform timeseries prediction tasks.

\end{abstract}

%
%
%
%
%

\section{Introduction}

In recent years there has been significant progress in developing neuromorphic hardware that complements the conventional computing architecture. Especially in-memory devices based on memristive or memtransistive elements are in focus as they can change their resistance based on the history of the applied voltage, resulting in a memory effect \cite{CHU22x,ZHO24}. These hysteresis effects  can be used either to  switch the resistance \cite{IEL18,MAT23a} or to exploit the temporal rate-dependent response \cite{DU17}. There are also efforts to realize optical switching with electro-optical memristor realizations \cite{EMB20,HU21,LIU22b}.

The nonlinearity and fading memory of memristors makes them suitable candidates for scalable and low power consuming physical reservoir computing (RC) \cite{JAN24a}. 
Various material systems have already been used for RC realizations based on the rate dependent response, including SiC electrolyte as switching layer \cite{GUO23c} , electron trap based W/HfO$_2$/TiN memristors \cite{JAN21}, or vertically stacked oxide based structures \cite{ZHO21, LU24, ZHO19a, PAR21c}. Combinations of volatile and nonvolatile memory have been investigated using organic electrochemical synaptic transist\cite{LIU24a} as well as van-der-Waals semiconductor-based $\alpha$-phase indium selenide devices with optically tunable nonlinearity \cite{LIU22b}.
In all those works the focus has mainly been on utilizing the dynamics of systems which exhibit rate-dependent hysteresis. However, there are also many devices which exhibit rate-independent hysteresis including magnetic domains used for RC in \cite{ZHO25a} and bistable organic electrochemical transistors \cite{BON24}. See also a recent review on the development of memristors based on different physical mechanism and their potential use for neuromorphic computing \cite{CHE23d}.

The switching properties of memristive devices that show hysteresis has been widely used to realize logic bit operations, in-memory computing and matrix multiplications using crosspoint memory architectures \cite{IEL18}. For this, many spatially distributed devices are used either in a 2D \cite{JO09} or in 3D \cite{KAU09} arrays. The goal in these setups is to realize digital operations like matrix multiplication in a fast and energy efficient way without the need for data traffic between computing and storage units. These arrays are also used to realize the layers in artificial neural networks using the fact that the connection weights can be trained via switching \cite{IEL18}.

In this work we have asked ourselves the simple question: Can hysteresis be used to perform nonlinear transforms, i.e. can we realize an analog computing scheme via utilizing only hysteresis between two linear curves realized with only one instantaneous switching device?  We do not include any delay or transient behavior for additional memory, as e.g. used in \cite{CAR24a}, and only utilize the fact that the system's resistance (the input-output relation) depends on its previous state. Since we use a very simple system to address this fundamental question, our results are not only applicable to memristors, but to any system exhibiting such hysteresis.


To address the computational potential of our simple hysteresis system, we insert time-multiplexed inputs into the system and generate the outputs via a weighted linear sum, similar to extreme learning machines \cite{HUA04} and time-mutliplexed reservoir computing \cite{HUE22, APP11,BRU13a}. We however, do not exploit the finite switching times and the corresponding variations in the nonlinear response due to transients and operate the system in a discrete time setting as an instantaneous switch with hysteresis. We investigate the nonlinear transform capabilities of the system \cite{DAM12,HUE22a}, the consistency of the input-response \cite{LYM19a, VER21, UCH04} and the influence of system noise on the performance. The system is essentially memoryless, however multiple studies have shown how delays in the input or output layers can compensate for this if needed \cite{MAR19a,JAU21a,JAU24,PAR24,JAU25}. Memory and dynamics can also be generated by coupling multiple nodes with hysteresis, however, in this work we focus on characterising and understanding the performance of a single system using only the rate-independent switching response.

The important questions that we want to address with our study are: Do we get sufficient nonlinear transform using only the instantaneous switch with hysteresis? Do we have enough consistency to realize a analog computer device despite the fact that the system response depends on the branch it is on? 

For the hysteresis curves we have drawn inspiration from the current-voltage characteristics of memristors \cite{IEL18,WAN18f, ABE24}. However, we simplified the input-output characteristic such  that the results can be generalised to many types of devices, i.e. we consider two linear branches, set two threshold for switching between them and set rules for the switching event depending on the branch the system is on. 
We wish to show that this simple system used in an RC setup has the capacitiy to perform time series prediction tasks. In that respect the goal is not a comparison with established RC realizations  but a comparison to a purely linear system, given the fact that our system is linear with the addition of hysteresis induced switching. 
The challenge for experimentalist is to see if this type of hysteresis can be implemented more easily/ more energy efficiently than other options, and to see if the performance is sufficient for the required task.





\section{Methods}

\subsection{Hysteresis Model}

\begin{figure}[t]
\centerline{\includegraphics[width=0.95\textwidth]{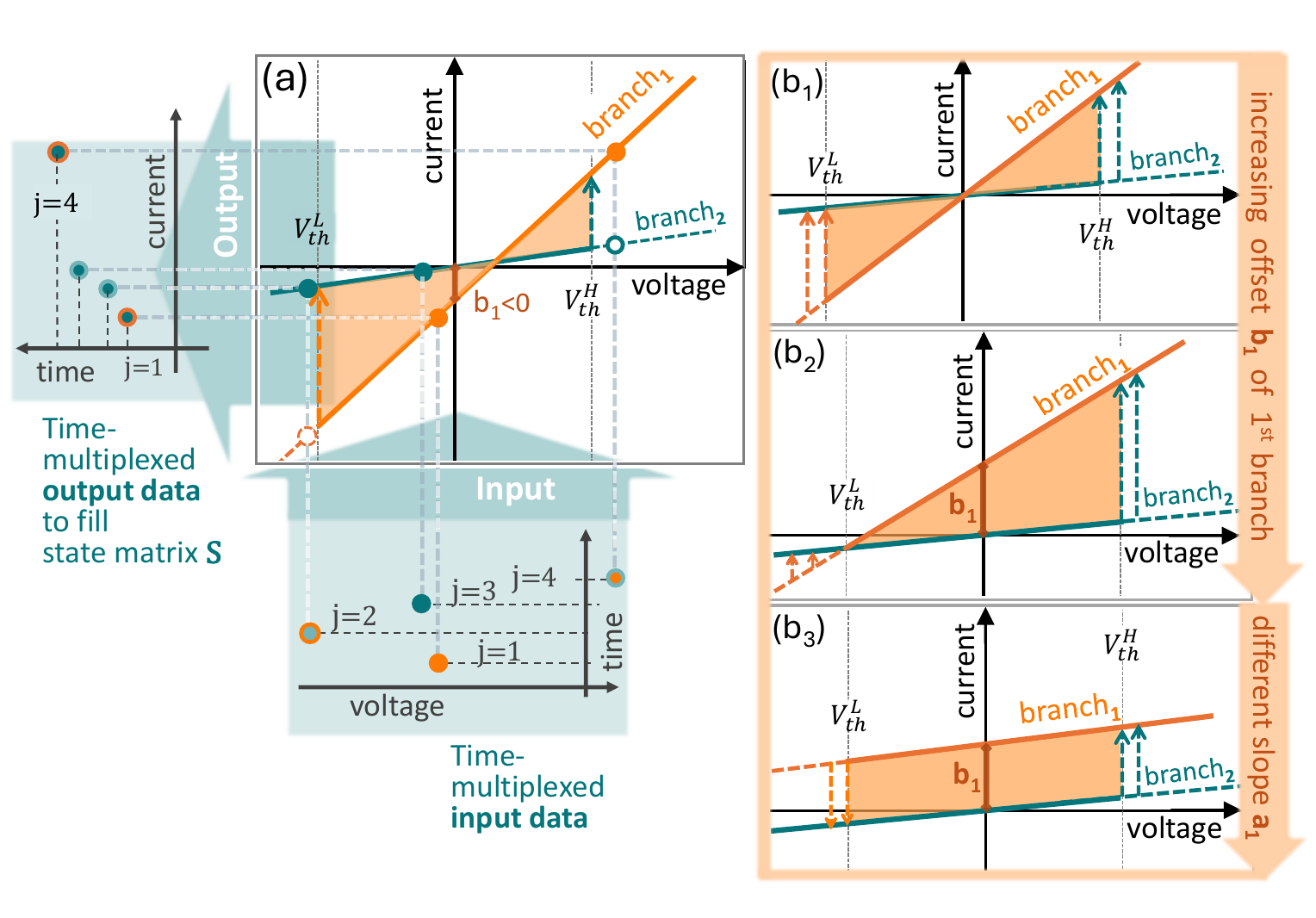}}
\caption{(a) Sketch of the Hysteresis model that is build via two linear branches (orange and green) and two switching points at $V_{th}^L$ and $V_{th}^H$ in the plane of input voltage and output current. Orange shadow indicates bistable region. The time-multiplexed input of the RC is fed in via voltage modulation while the model response is collected as time-dependent current (time indexed with $j$). ($b_1$)-($b_3$) indicate how the shape of the hysteresis changes when the offset $b_1$ of the orange branch$_1$ is increased and when the slope $a_1$ of the first branch is changed to equal $a_2$ ($b_3$).}
\label{fig1}
\end{figure}

Our hysteresis model is motivated by the voltage-current characteristics of memristors \cite{IEL18, WAN18f}. We will therefore refer to the input variable as voltage and output variable as current. However, we wish to emphasize that this model could represent any system which displays hysteresis of a similar form. Furthermore, compared with the characteristics of many existing memristor devices \cite{IEL18}, our model is overly simplified. This is intentional as we aim to understand how hysteresis can be utilised for computations independently from additional nonlinearities in the input-output behavior.

The hysteresis consists of two linear branches with two switching points, as illustrated in Fig.~\ref{fig1}. The branch switching points are set to $V^L_{th}=-1$ and $V^H_{th}=1$. The output current depends on the voltage (input) and the branch the system is currently on, as described by Algorithm 1 and visualized in Fig.\ref{fig1}a. Parameters are the offsets, $b_{1,2}$, and slopes, $a_{1,2}$, of the branches, and the switching points $V_{th}^{L,H}$.

\begin{algorithm}
\caption{Hysteresis Model}\label{model}
\begin{algorithmic}
\If{(Branch$_1$ is $\bf{true}$ \textbf{and} $V\geq V^L_{th}$) $\bf{or}$ (Branch$_2$ is $\bf{true}$ \textbf{and} $V > V^H_{th}$)}
    \State $I = a_1V+b_1$
    \State Branch$_1=\bf{true}$
    \State Branch$_2=\bf{false}$ 
\ElsIf{(Branch$_1$ is $\bf{true}$ \textbf{and} $V < V^L_{th}$) $\bf{or}$ (Branch$_2$ is $\bf{true}$ \textbf{and} $V \leq V^H_{th}$)}
    \State $I = a_2V$
    \State Branch$_1=\bf{false}$
    \State Branch$_2=\bf{true}$
\EndIf
\end{algorithmic}
\end{algorithm}

\subsection{Time-multiplexing and regression}

To use the hysteresis between the two linear branches for computing, we use an approach similar to reservoir computing and extreme learning machines \cite{APP11,BRU13a,HUA04,HUE22,ORT15}. We feed randomly masked time-multiplexed data into the hysteresis model and linearly combine the responses to approximate the desired output. The linear output weights are trained via ridge regression as detaled below. 

\subsubsection*{Masking and time-multiplexing\\}

A given task-dependent input sequence ${\bf u}=[u_1,...,u_K]$ is masked and time-multiplexed by multiplying each $u_k \in {\bf u}$ by a sequence of mask values $[m_1,...,m_M]$. The final masked and time-multiplexed input sequence of length $K\times M$ is given by:
\begin{equation}
    {\bf v}=[u_1m_1,...,u_1m_M,...,u_Km_1,...,u_Km_M].
\end{equation}
If there are multiple input sequences, then we mask them individually and then add them together:
\begin{equation}
    {\bf v}=\sum_i[u^i_1m^i_1,...,u^i_1m^i_M,...,u^i_Km^i_1,...,u^i_Km^i_M],
\end{equation}
i.e. ${\bf v}=[V_1,...,V_L]$ with $L=K\times M$, and $V_j=\sum_i u^i_km^i_m$ with $j=(k-1)M+m$.
The input voltages to the hysteresis model described by Algorithm 1 are then given by $V=GV_j$, where $G$ is an input scaling parameter and all $GV_j$ are fed in sequentially (see Fig.\ref{fig1}a).
The mask values are drawn from a uniform distribution between -1 and 1.

\subsubsection*{Training\\}

All responses of the hysteresis system to input $u_k$ are collected in a vector $[I_1,...,I_M]$. Appending a bias term of one to the end of this vector, we have the feature vector ${\bf s_k}=[I_1,...,I_M,1]$. The output of the system is then given by
\begin{equation}
    \hat{y}_k={\bf s_k} {\bf w},
\end{equation}
where ${\bf w}=[w_1,...,w_{M+1}]^{\textrm{T}}$ is a vector of weights. The weights are trained by minimising the squared Eucledian distance between $\hat{y}_k$ and the target outcome $y_k$:
\begin{equation}
    \min_{\bf w} \left( || {\bf S} {\bf w}-{\bf y}||^2 +\lambda ||{\bf w}||^2\right),
\end{equation}
where ${\bf S}$ is an $K\times (M+1)$ matrix comprised of the feature vectors ${\bf s_k}$ corresponding to the input sequence ${\bf u}$, ${\bf y}=[y_1,...,y_K]$ is the vector of target outputs and $\lambda$ is a regularisation parameter. The solution to this minimisation problem is given by 
\begin{equation}
    {\bf w}=\left( {\bf S}^\textrm{T}{\bf S}+\lambda {\bf I}\right)^{-1}{\bf S}^\textrm{T}{\bf y},
\end{equation}
where ${\bf I}$ is the $(M+1)\times (M+1)$ identity matrix. 

\subsection{Tasks and performance measures}
\label{lorenz}

We evaluate the computational performance of the time-multiplexed hysteresis system using different measures, these are the nonlinear transform capacities and two timeseries prediction tasks (Lorenz and Mackey-Glass) which are explained in the following:

\subsubsection*{Nonlinear transform capacity\\}

We define the nonlinear transform capacity as a subset of the information processing capacity (IPC) introduced in \cite{DAM12}. The IPC is a measure of nonlinear transform and memory capabilities, mainly used in the context of reservoir computing. Since the hysteresis system studied here has no memory beyond which branch the system is currently on, we are only interested in the nonlinear transform aspect of the IPC. We therefore only use the IPCs corresponding to the current input (all IPCs involving past inputs are zero due to the lack of memory of the system). To calculate the nonlinear transform capacities, the input sequence ${\bf u}$ is a sequence of random numbers drawn from a uniform distribution on the interval $[-1,1]$ and the target sequences are Legendre polynomials as a function of the input values, as described in \cite{HUE22a} using only the current input. The capacities are then given by 
\begin{equation}
    C_i=1-\textrm{NMSE}_i,
\end{equation}
where $\textrm{NMSE}_i$ is the normalised mean square error (NMSE) of the approximation of the i$^\textrm{th}$ order Legendre polynomial. The NMSE is defined as square of the normalised root mean square error given by Eq.~(\ref{NRMSE}).

\subsubsection*{Lorenz 63 timeseries prediction\\}

The Lorenz 63 system \cite{LOR63} is given by 
\begin{eqnarray}
\frac{dx}{dt}=c_1y-c_1x,\quad
\frac{dy}{dt}=x(c_2-z)-y, \quad \textrm{and} \quad
\frac{dz}{dt}=xy-c_3z.\label{Lorenz}
\end{eqnarray}
With $c_1=10$, $c_2=28$ and $c_3=8/3$ this system exhibits chaotic dynamics. We generate the Lorenz timeseries using Runge-Kutta fourth order numerical integration with a time step of $h=10^{-3}$. We use all three variables, sampled with a step size of $\Delta t=0.02$ and rescaled to the range $[0,1)$, as the hysteresis input sequences $u^1_k=X(k\Delta t)$, $u^2_k=Y(k\Delta t)$ and $u^3_k=Z(k\Delta t)$ ($X$, $Y$, $Z$ indicate the rescaled versions of $x$, $y$, $z$). In the training phase the targets are given by $ y^1_k=X((k+1)\Delta t)$, $y^2_k=Y((k+1)\Delta t)$ and $y^3_k=Z((k+1)\Delta t)$.

We have chosen this variant of the Lorenz task as no memory is needed to make the prediction when all three variables are provided as input \cite{JAU24a}.

\subsubsection*{Mackey-Glass timeseries prediction\\}

The Mackey-Glass delay-differential equation is given by \cite{MAC77}
\begin{equation}\label{eqMG}
\frac{dx}{dt}=\beta \frac{x \left( t-\tau_M \right) }{1+x \left( t-\tau_M \right)^n}-\gamma x.
\end{equation}
Using the standard parameters, $\tau_M=17$, $n=10$, $\beta=0.2$ and $\gamma=0.1$, this system exhibits chaotic dynamics. In this study we use $12$-step-ahead prediction of this system with an input discretisation of $\Delta t=1$ as our task. To create the input sequence we generate a timeseries by numerically integrating Eq.~(\ref{eqMG}) using a Runge-Kutta fourth order method with Hermitian interpolation for the mid-steps of the delayed terms and with a time step of $h=10^{-2}$. This timeseries is then sampled with a time step of $\Delta t=1$. The input to the hysteresis system is then given by the sequence $u_k=X(k\Delta t)$, where $X(t)$ is $x(t)$ rescaled to the range $[0,1)$. The corresponding target sequence is given by $y_k=X((k+12)\Delta t)$. To compensate for the lack of memory of the time-multiplexed hysteresis system we use the delayed-state-matrix concatenation described in \cite{MAR19a, JAU24,JAU25}, using a delay of 9 steps, i.e. each row of the final state matrices used for training and prediction includes the system responses to the current input and the input from 9 steps in the past.

\subsubsection*{Error Measure\\}

The performance of the timeseries prediciton tasks is quantified using the normalised root mean squared error (NRMSE), defined as
\begin{equation} \label{NRMSE}
 \textrm{NRMSE}=\sqrt{\frac{\sum_{k=1}^{K_T}\left( y_{k}-\hat{y}_{k}\right)^2}{K_{Te} \textrm{var}\left( {\bf y}\right)}},
\end{equation}
where $y_k$ are the target values, $\hat{y}_{k}$ are the outputs produced by the time-multiplexed hysteresis system, $K_{Te}$ is the number of testing steps (i.e. the length of the vector ${\bf y}$) and $\textrm{var}\left( {\bf y} \right)$ is the variance of the target sequence.

\subsection{Simulation parameters}\label{sec:simpara}

For all simulations the input sequences are divided into four sections; buffer ($K_B=100$ steps), training ($K_{Tr}=10000$ steps), buffer ($K_B=100$ steps), testing ($K_{Te}=5000$ steps). Here the steps refer to the steps $k$ of the task-dependent input sequences ${\bf u}=[u_1,...,u_K]$. The regularisation parameter in the ridge regression is kept constant at $\lambda=5\times 10^{-6}$ for all simulations. Unless stated otherwise all results are the median of 200 realisations of the system with different randomly drawn input masks and the variability is given by the median absolute deviation (MAD). 

\section{Results and discussion}

\begin{figure}[t]
\centerline{\includegraphics[width=1\textwidth]{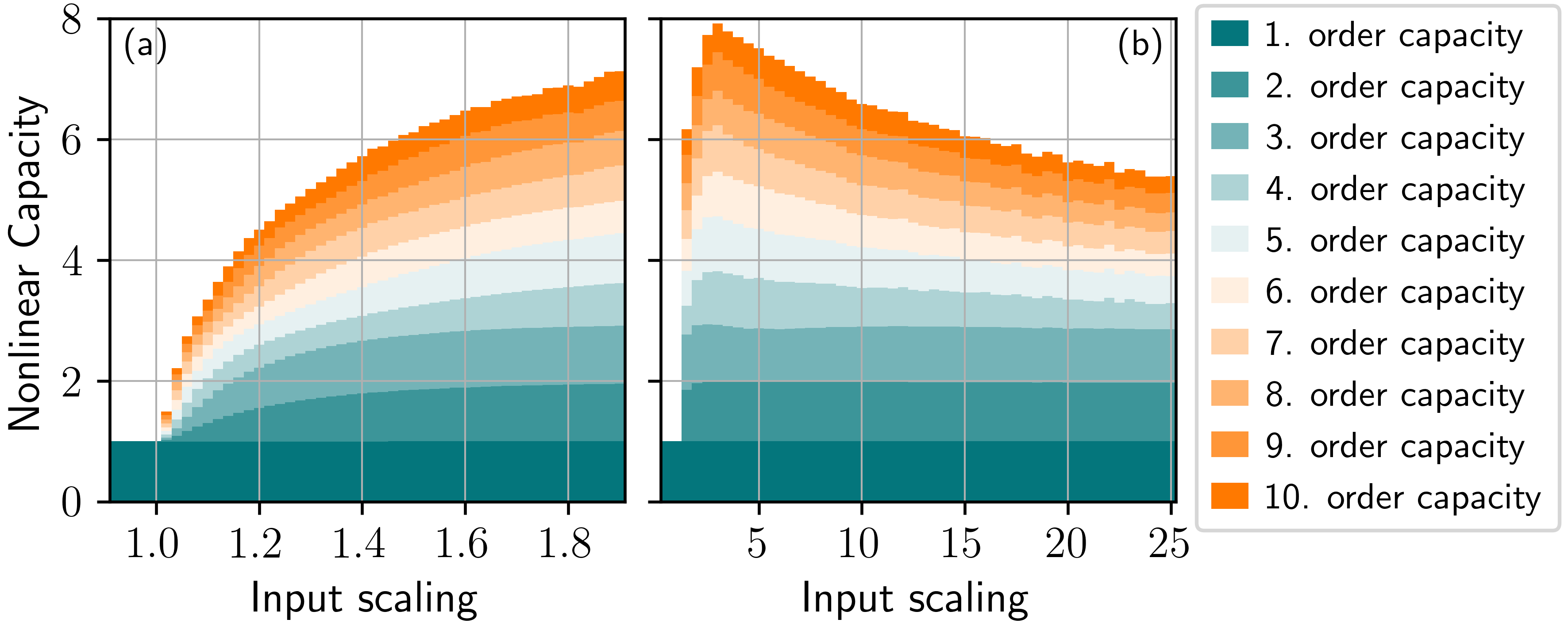}}
\caption{Median nonlinear capacities $C_i$ of orders $i\in [1,10]$ as a function of the input scaling $G$ for low (a) and high (b) ranges of $G$. Parameters: $a_1=0.6$, $a_2=0.5$, $b_1=0.1$, $M=100$, and as in Section~\ref{sec:simpara}.}
\label{fig2}
\end{figure}

We investigate the performance of the time-multiplexed hysteresis model as a function of four parameters; the input scaling $G$, the number of features (number of mask values) $M$, the slope of branch one $a_1$ and the offset of branch one $b_1$. Without switching between the branches, the system is purely linear. Therefore, if $G \leq 1$, the input never drives the system beyond the switching points at $V=\pm1$ and the nonlinear transform capacities above first order are all zero, as seen in Fig.~\ref{fig2}a where only dark green appears for $G<1$. For the figure, capacities up to 10$^\textrm{th}$ order were calculated and plotted with colors from green to orange with bright orange being the 10$^\textrm{th}$ order transform capacity. As soon as the switching between branches can occur above $G=1$, the system has the capacity to perform nonlinear transforms. For $G$ just above one, mask-dependent switching events are rare, resulting in low nonlinear transform capacities. As $G$ is further increased the lower order capacities reach one (representing perfect reconstruction of these nonlinear transforms). For  large input scaling beyond $G=3$ the higher order capacities decrease again (see Fig.~\ref{fig2}b). This is because an increasing proportion of the masked inputs are outside the switching range, meaning that the influence of the hysteresis is reduced. 

\begin{figure}[t]
\centerline{\includegraphics[width=0.5\textwidth]{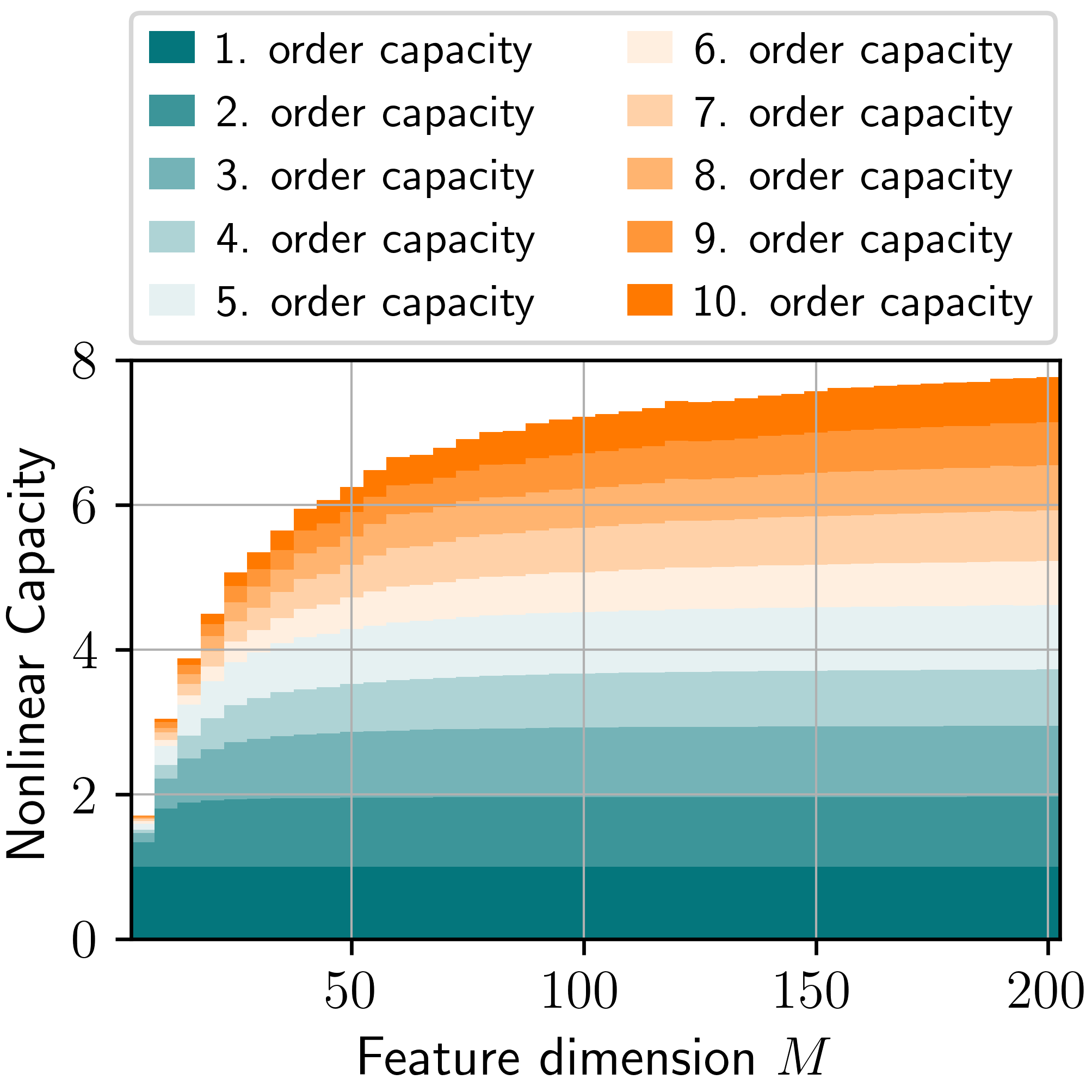}}
\caption{Median nonlinear capacities $C_i$ of orders $i\in [1,10]$ as a function of the feature dimension $M$. Parameters: $a_1=0.6$, $a_2=0.5$, $b_1=0.1$, $G=2$, and as in Section~\ref{sec:simpara}.}
\label{fig3}
\end{figure}

The nonlinear transform capacities also depend on the feature dimension, i.e. number of masked inputs $M$, as depicted in Fig.~\ref{fig3} for a gain of $G=2$. It is noted that the higher $M$, the more data have to be fed in sequentially per input which leads to slower processing rates. With increasing $M$ all capacities which have not reached one (full capacity) increase. As typical in time-multiplexed reservoir computing there is trade-off between performance and the processing rate \cite{BRU18a,GOL23,MUE24}.

\begin{figure}[t]
\centerline{\includegraphics[width=1\textwidth]{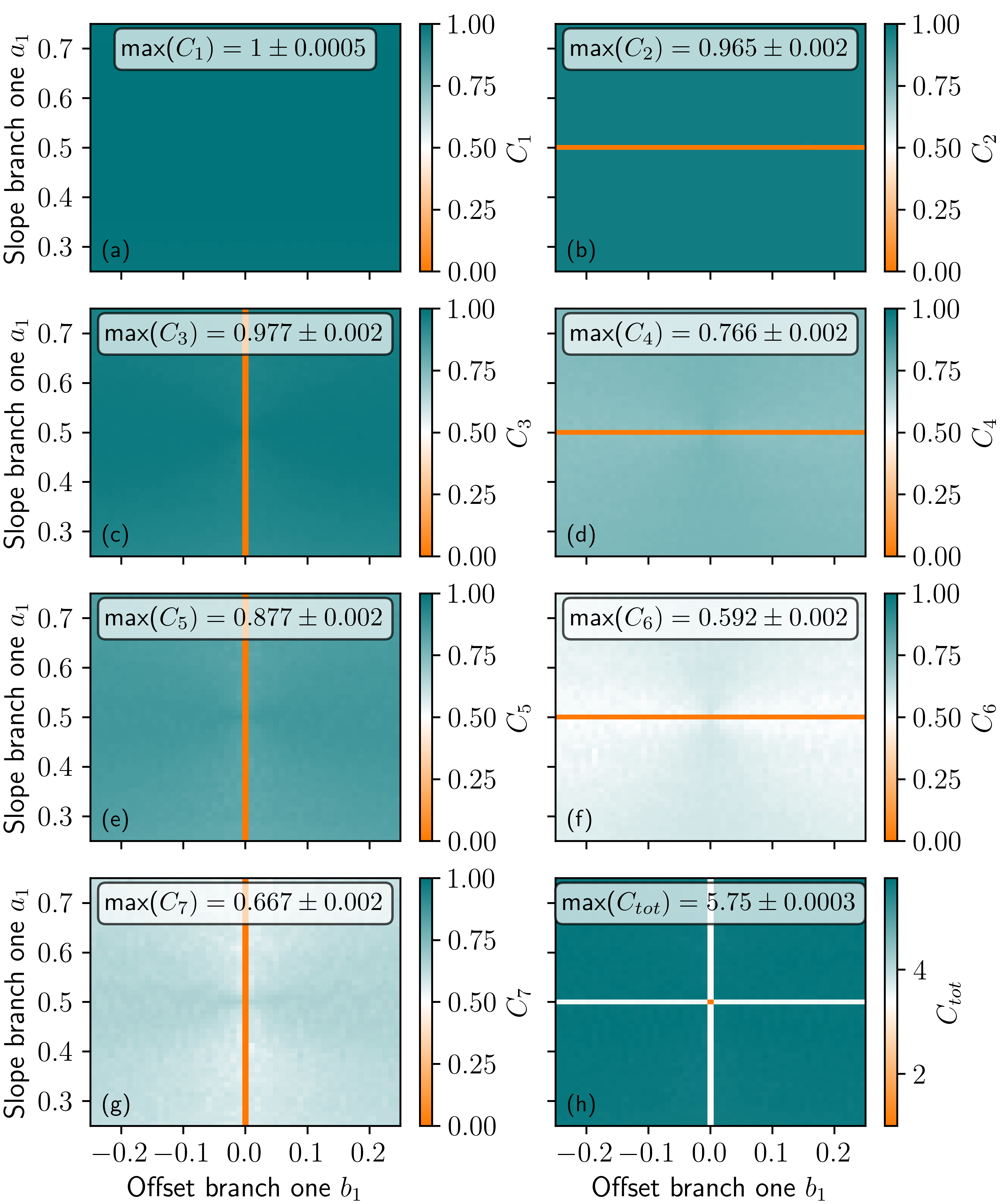}}
\caption{Median nonlinear capacities $C_i$ of orders $i\in [1,7]$ (a)-(g) as a function of the slope $a_1$ and offset $b_1$ of branch one. (h) Sum over the capacities $C_1$ to $C_7$. The maximum capacity for each subplot is indicated along with the correspond MAD. Parameters: $a_2=0.5$, $G=2$, $M=100$, and as in Section~\ref{sec:simpara}.}
\label{fig4}
\end{figure}

In Fig.~\ref{fig2} and Fig.~\ref{fig3} fixed values for the slopes and offsets of the two branches were used. As sketched in Fig.\ref{fig1}b both parameters can drastically change the shape of the hysteresis curve. In Fig.~\ref{fig4} we investigate the behavior of the nonlinear capacities as a function of the slope $a_1$ and offset $b_1$ of branch one, and find features corresponding to certain symmetries between the branches. For the even nonlinear transforms plotted in Fig.\ref{fig4}b,d,f
the capacities are zero (as seen by the horizontal orange lines) when the slopes of the two branches are both equal at $m_1=m_2=0.5$ (corresponding hysteresis is sketched in Fig.~\ref{fig1}b$_3$). Instead, for the odd nonlinear transform capacities plotted in Fig.\ref{fig4}c,e,g the values are zero (vertical orange lines) when the offsets of both branches are zero, $b_1=b_2=0$ (as sketched in Fig.~\ref{fig1}b$_1$). Note the offset of branch two is zero throughout this study. Aside from these zero capacity regions, there is very little dependence on $a_1$ and $b_1$. The linear capacity (Fig.~\ref{fig4}a) is independent of $a_1$ and $b_1$ due to the linearity of the branches. 

\begin{figure}[t]
\centerline{\includegraphics[width=1\textwidth]{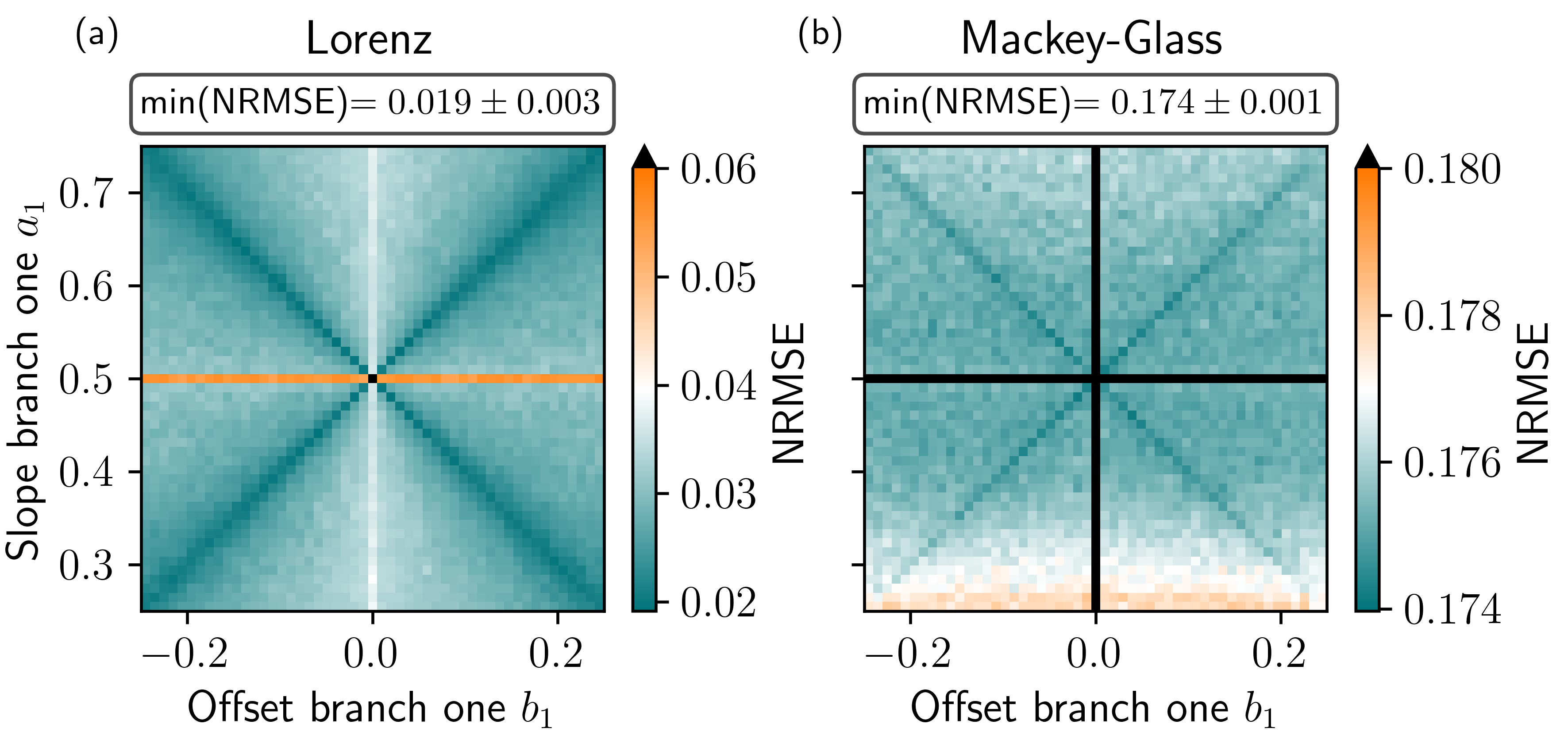}}
\caption{Median NRMSE for (a) the z-component of the Lorenz task and (b) the Mackey-Glass task, as a function of the slope $a_1$ and offset $b_1$ of branch one. The minimum NRMSE for each task is indicated along with the correspond MAD. Parameters: (a) $G=2$, (b) $G=4$, $a_2=0.5$, $M=100$, $K_B=10000$, remaining parameters as in Section~\ref{sec:simpara}.}
\label{fig5}
\end{figure}

To test the performance of the time-multiplexed hysteresis system on actual tasks, we have chosen two timeseries prediction tasks, the results of which are shown in Fig.~\ref{fig5}. For the Lorenz task (see details in Section \ref{lorenz}), the NRMSE for predicting the z-component is shown in Fig.~\ref{fig5}a. From the equations for the Lorenz system (Eq.~(\ref{Lorenz})) it is clear that this task requires various nonlinear transforms of the input data. Therefore, we see features relating to the nonlinear transforms in the task performance, i.e. for all $a_1=0.5$ and all $b_1=0$, poor prediction performance is found corresponding to the zero values for the even and odd capacities in Fig.~\ref{fig4}b-g. Additionally, we find regions of good performance along the diagonal lines corresponding to $|a_1-a_2|=|b_1|$. For the system parameter fulfilling $|a_1-a_2|=|b_1|$ the crossing point of branch one and branch two is at one of the switching points $V=\pm 1$, an example of which is depicted in Fig.~\ref{fig1}b$_2$. It is unclear why this is beneficial for the performance. Overall a substantial performance improvement is achieved compared with a purely linear system. The linear system corresponds to the point right in the middle of Fig.~\ref{fig5}a where $a_1=0.5$ and $b_1=0$, as there both branches are identical and solely linear transformations occur. For a purely linear system the NRMSE for predicting the z-component is 0.182 if a feature dimension $M\geq3$ is chosen (note that we have not stated a MAD value for the NRMSE of the linear system, as in this case different masks do not influence the performance). For the purely linear system the performance saturates when $M$ is equal to the input dimension (which is 3 for the Lorenz task) because no additional linearly independent information can be added with more masked inputs.

For the Mackey-Glass task shown in Fig.~\ref{fig5}b, the same general features as in the Lorenz case, with best performance along the diagonals and worse performance along the axis $b_1=0$ and $a_1=0.5$, can be observed. The overall performance is however much worse, as this task requires more memory \cite{JAU24}. However, here we also find a substantial improvement compared with the linear system in the center of the 2D plot which yields an NRMSE of 0.600 for $M\geq2$ for Mackey-Glass 12-step ahead prediction ($M\geq2$ because here the input dimension is two as there is one input and state matrix concatenation with one delayed system response).

\begin{figure}[t]
\centerline{\includegraphics[width=1\textwidth]{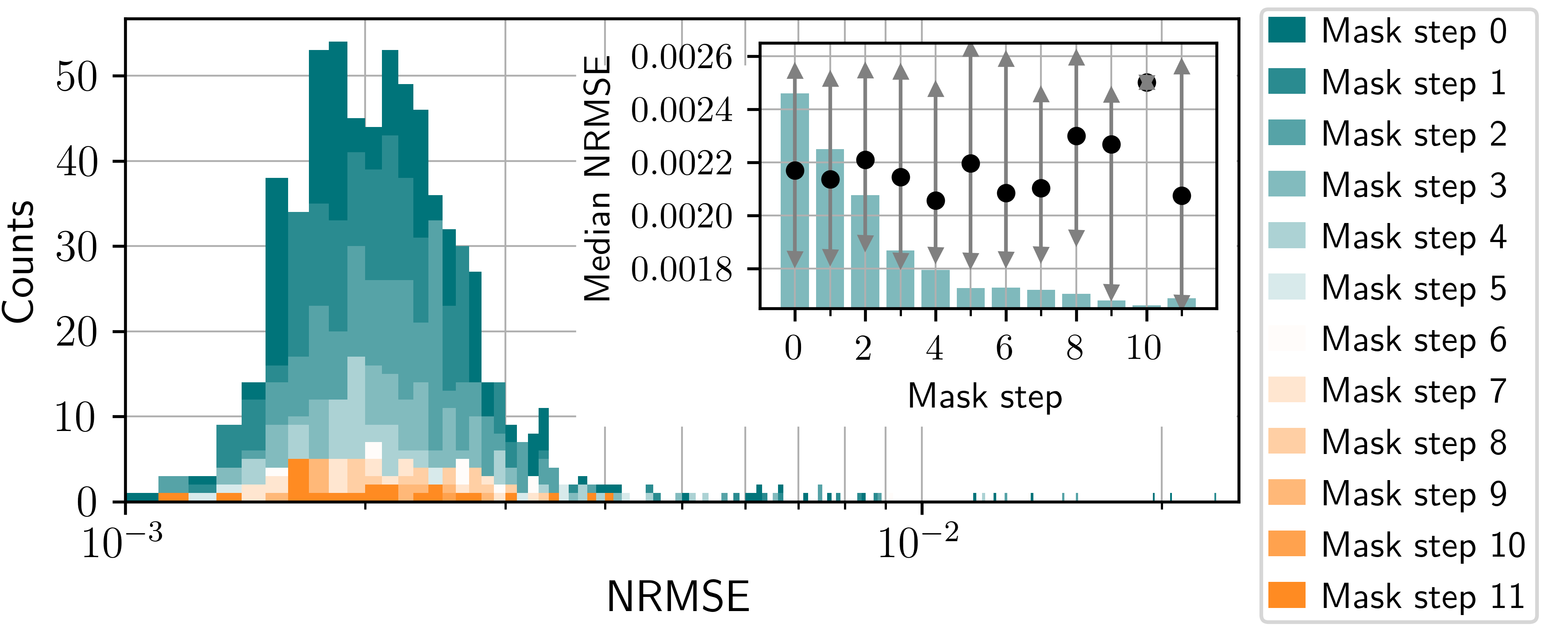}}
\caption{Histograms of the NRMSE of the x-component of the Lorenz task categorized by the median first-threshold mask step (different colors) for 2000 randomly drawn input masks. The inset gives the median NRMSE for each category (black circles), with the arrows indicating the ranges from the 25$^{\textrm{th}}$ to the 75$^{\textrm{th}}$ percentiles. The green bars in the inset indicate the relative number of samples in each category, with the highest being 656 for mask step zero and the lowest being 1 for mask step ten. Parameters: $a_1=0.6$, $a_2=0.5$, $b_1=0.1$, $G=2$, $M=100$,  $K_B=10000$, mask realisation$=2000$, remaining parameters as in Section~\ref{sec:simpara}.}
\label{fig6}
\end{figure}

For the time-multiplexed hysteresis system to reliably compute tasks, it is important that similar inputs lead to similar responses (consistency property). Since the response of the memristive-inspired node depends on the current branch-state of the system, two identical masked inputs can lead to different response sequences if the system starts on opposite branches.  Due to the multiplexing, one task input-value is transformed into a sequence of masked inputs which are fed into the hysteresis system. As soon as the masked input first surpasses the switching threshold of $V=\pm1$, the remaining responses for the mask sequence are identical for the two cases, i.e. the remaining responses are independent of the starting branch. To gain an understanding, if this "inconsistency" has an impact on the performance, we have calculated the median mask step at which the switching threshold $V=\pm1$ is first surpassed over the testing sequence of the Lorenz task, i.e. for each step in the Lorenz timeseries $k$ we have calculated the masked input $V_j=\sum_i^3 u^i_km^i_m$ ($j=(k-1)M+m$) and recorded the first $m$ for which $|V_j|\geq 1$, and compared this with the performance for 2000 randomly drawn masks. Figure~\ref{fig6} shows histograms of the NRMSE of the Lorenz x-component for the mask realisations sorted according to their median first-threshold mask step. These histograms, along with the corresponding median NRMSEs (inset of Fig.~\ref{fig6}), show no clear performance differences between masks which result in earlier or later switches. Thus, the inconsistency which can arise due to the branch switching does not appear to impact the performance.

\begin{figure}[t]
\centerline{\includegraphics[width=0.55\textwidth]{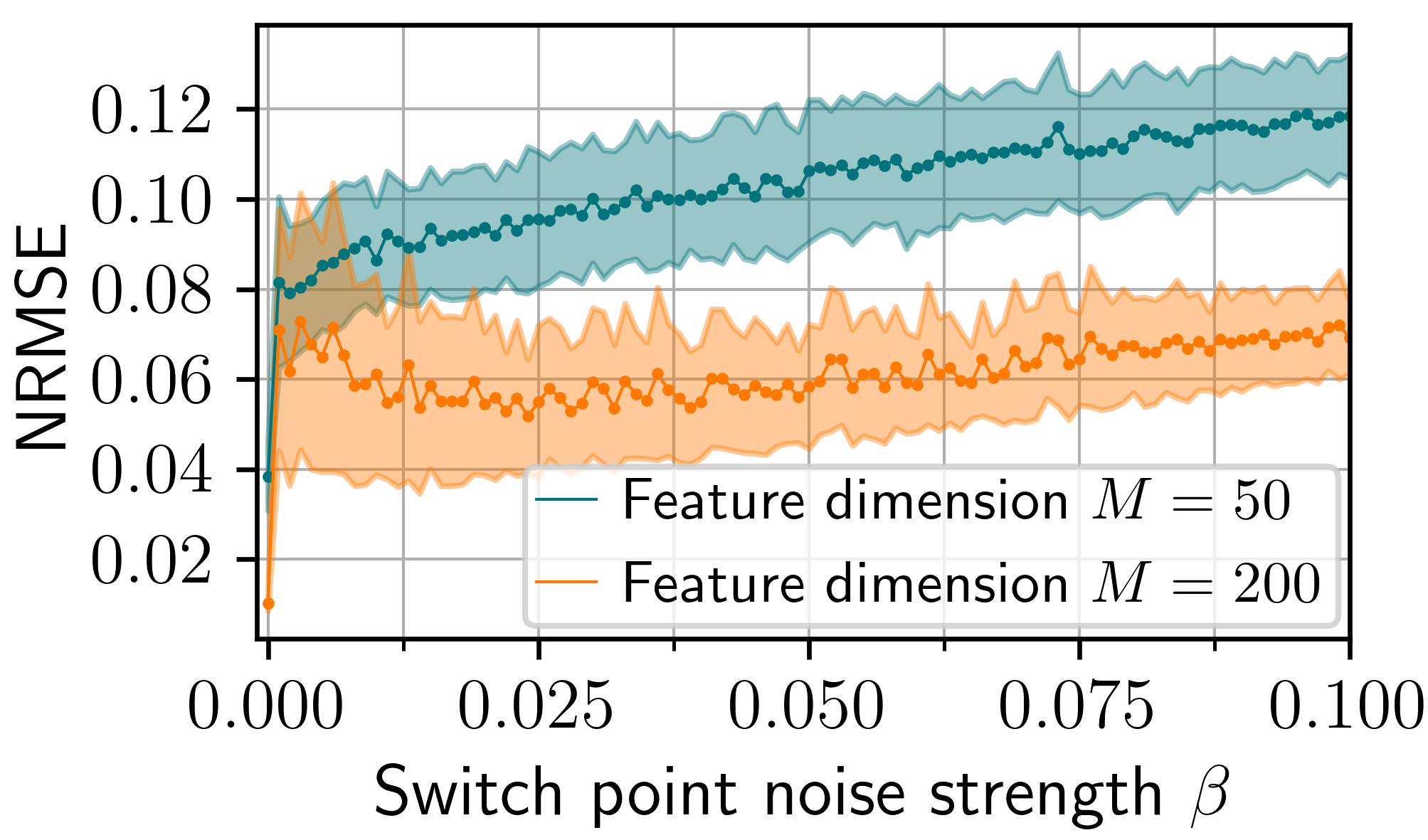}}
\caption{Median NRMSE of the z-component of the Lorenz task as a function of the noise strength $\beta$ for $M=50$ and $M=200$. The shaded regions indicted the MAD. Parameters: $a_2=0.5$, $G=2$, $K_B=10000$, remaining parameters as in Section~\ref{sec:simpara}.}
\label{fig7}
\end{figure}

Finally, we address the influence of noise on the system. Our hysteresis model is motivated by memristors, which can show variability in the switching thresholds \cite{WAN18f}. We deem this to be the most important type of noise to simulate, as differences in switching can influence the responses of subsequence mask steps whereas noise in the branch properties would only influence individual responses. To simulate noise in the switching thresholds, we have added a noise term to the thresholds; $V^L_{th}\rightarrow V^L_{th} +\beta \xi (j)$ and $V^H_{th}\rightarrow V^H_{th} +\beta \xi (j)$, where $\xi (j)$ is drawn from a Gaussian distribution of zero mean and standard deviation one at each time-multiplexed iteration step $j$. The factor $\beta$ scales the noise strength. The performance for predicting the z-component of the Lorenz task is shown as a function of the noise strength $\beta$ in Fig.~\ref{fig7}. Even with small noise strengths, there is a strong degradation in the performance, as seen by the sudden increase in the error compared with zero noise. However, over the entire $\beta$ range an improvement over the linear system (with an NRMSE of 0.182) is maintained.

\section{Discussion and conclusions}

We have shown that hysteresis between two linear branches of a single system can be utilized to perform nonlinear analog computation tasks es for example time series prediction. In contrast to rate-dependent hysteresis which has already been widely used in a time-multiplexed reservoir computing scheme, we solely exploit the rate-independent hysteresis property via implementing instantaneous and state-dependent switching between the branches. This work adds to the fundamental understanding of how physical devices exhibiting hysteresis can be utilized for analog computations. 

We show that the single investigated memristor-inspired device can be used for time series prediction tasks and evaluate the nonlinear transformations that can be realized. The prediction performance exceeds that of a purely linear system even if noise in the switching characteristics is considered. Surprisingly, inconsistencies that exist due to the state-dependent response of the device, do not have a considerable impact on the performance. 

We note that for the case that the hysteresis branches are themselves nonlinear curves, improved performance can be expected and the scaling behavior with the feature dimension will be more favorable. We therefore hope that this work will inspire new approaches to building fast and efficient analog computing devices as there may be existing physical devices or electronic implementations where such a way of using the hysteresis will contribute beneficially to the performance. 
Open questions include how such systems with rate-independent hysteresis would behave when coupled and operated in a spatially multiplexed setup instead of the pure time-multiplexing discussed here.

\section*{Acknowledgements}

This work was made possible by funding from the Carl-Zeiss-Stiftung.

\section*{References}
\providecommand{\newblock}{}

\end{document}